\documentclass[12 pt,twoside,aps,prd,amsmath,amssymb,
tightenlines,showpacs,showkeys,eqsecnum]{revtex4}

\usepackage{graphicx}
\usepackage{mathptmx}
\usepackage{bm}
\newcommand {\ve}{\varepsilon}

\newcommand {\cD}{\cal D}

\newcommand {\G}{\Gamma}
\newcommand {\bp}{\bar \psi}

\def\myfigure #1#2#3#4
{\begin{figure}[ht]\begin{center}
\includegraphics[width=#2 \textwidth]{#1.eps}
\parbox[t]{#4\textwidth}{\caption{#3}\label{#1}}
\end{center}\end{figure}}

\def \myfigures #1#2#3#4#5#6#7#8
{\begin{figure}[ht]
    \begin{center}
        \includegraphics[width=#2 \textwidth]{#1.eps}
        \hfill
        \includegraphics[width=#6 \textwidth]{#5.eps}
        \parbox[t]{#4\textwidth}{\caption {#3}\label{#1}}
        \hfill
        \parbox[t]{#8\textwidth}{\caption {#7}\label{#5}}
    \end{center}
\end{figure} }

\begin{document}
\baselineskip -24pt
\title{Spinor model of a perfect fluid and their applications in Bianchi type-I and FRW models}
\author{Bijan Saha}
\affiliation{Laboratory of Information Technologies\\
Joint Institute for Nuclear Research, Dubna\\
141980 Dubna, Moscow region, Russia} \email{bijan@jinr.ru}
\homepage{http://wwwinfo.jinr.ru/~bijan/}

\begin{abstract}

Different characteristic of matter influencing the evolution of the
Universe has been simulated by means of a nonlinear spinor field.
Exploiting the spinor description of perfect fluid and dark energy
evolution of the Universe given by an anisotropic Bianchi type-I
(BI) or isotropic Friedmann-Robertson-Walker (FRW) one has been
studied.

\end{abstract}



\maketitle

\bigskip

\section{Introduction}

Detection and further experimental reconfirmation of current cosmic
acceleration pose to cosmology a fundamental task of identifying and
revealing the cause of such phenomenon. In light of this
cosmological models with spinor field being the source of
gravitational field attracts widespread interest in recent time
\cite{henprd,sahaprd,greene,SBprd04,kremer1,ECAA06,sahaprd06,BVI,kremer2}.
It was shown that a suitable choice of spinor field nonlinearity\\
(i) accelerates the isotropization process
\cite{sahaprd,SBprd04,ECAA06};\\ (ii) gives rise to a
singularity-free Universe \cite{sahaprd,SBprd04,ECAA06,BVI} and \\
(iii) generates late time acceleration
\cite{kremer1,sahaprd06,kremer2}.

Question that naturally pops up is, if the spinor field can redraw
the picture of evolution caused by perfect fluid and dark energy, is
it possible to simulate perfect fluid and dark energy by means of a
spinor field? Affirmative answer to this question was given in the a
number of papers \cite{shikin,spinpf0,spinpf1}. In those papers the
authors have shown that different types of perfect fluid and dark
energy can be described by nonlinear spinor field. In \cite{spinpf0}
we used two types of nonlinearity, one occurs as a result of
self-action and the other resulted from the interaction between the
spinor and scalar field. It was shown that the case with induced
nonlinearity is the partial one and can be derived from the case
with self-action. In \cite{spinpf1} we give the description of
generalized Chaplygin gas and  modified quintessence \cite{pfdenr}
in terms of spinor field and study the evolution of the Universe
filled with nonlinear spinor field within the scope of a Bianchi
type-I cosmological model. The purpose of this paper is to extend
that study within the scope of a isotropic and homogeneous FRW
cosmological model.

\section{Simulation of perfect fluid with nonlinear spinor field}

In this section we simulate different types of perfect fluid and
dark energy by means of a nonlinear spinor field.

\subsection{Spinor field Lagrangian}

Let us first write the spinor field Lagrangian \cite{sahaprd}:
\begin{equation}
L_{\rm sp} = \frac{i}{2} \biggl[\bp \gamma^{\mu} \nabla_{\mu} \psi-
\nabla_{\mu} \bar \psi \gamma^{\mu} \psi \biggr] - m\bp \psi + F,
\label{lspin}
\end{equation}
where the nonlinear term $F$ describes the self-action of a spinor
field and can be presented as some arbitrary functions of invariant
generated from the real bilinear forms of a spinor field. For
simplicity we consider the case when $F = F(S)$ with $S = \bp \psi$.
Here $\nabla_\mu$ is the covariant derivative of spinor field:
\begin{equation}
\nabla_\mu \psi = \frac{\partial \psi}{\partial x^\mu} -\G_\mu \psi,
\quad \nabla_\mu \bp = \frac{\partial \bp}{\partial x^\mu} + \bp
\G_\mu, \label{covder}
\end{equation}
with $\G_\mu$ being the spinor affine connection. Varying
\eqref{lspin} with respect to $\bp (\psi)$ one finds the spinor
field equations:
\begin{subequations}
\label{speq}
\begin{eqnarray}
i\gamma^\mu \nabla_\mu \psi - m \psi + \frac{dF}{dS} \psi &=&0, \label{speq1} \\
i \nabla_\mu \bp \gamma^\mu +  m \bp - \frac{dF}{dS} \bp  &=& 0,
\label{speq2}
\end{eqnarray}
\end{subequations}
Variation of \eqref{lspin} with respect to metric function gives
energy-momentum tensor for the spinor field
\begin{equation}
T_{\mu}^{\rho}=\frac{i}{4} g^{\rho\nu} \biggl(\bp \gamma_\mu
\nabla_\nu \psi + \bp \gamma_\nu \nabla_\mu \psi - \nabla_\mu \bar
\psi \gamma_\nu \psi - \nabla_\nu \bp \gamma_\mu \psi \biggr) \,-
\delta_{\mu}^{\rho} L_{\rm sp} \label{temsp}
\end{equation}
where $L_{\rm sp}$ in account of spinor field equations
\eqref{speq1} and \eqref{speq2} takes the form
\begin{equation}
L_{\rm sp} = - S \frac{dF}{dS} + F(S). \label{lsp}
\end{equation}

We consider the case when the spinor field depends on $t$ only. In
this case for the components of energy-momentum tensor we find
\begin{subequations}
\begin{eqnarray}
T_0^0 &=& mS - F, \label{t00s}\\
T_1^1 = T_2^2 = T_3^3 &=& S \frac{dF}{dS} - F. \label{t11s}
\end{eqnarray}
\end{subequations}
A detailed study of nonlinear spinor field was carried out in
\cite{sahaprd,SBprd04,ECAA06}. In what follows, exploiting the
equation of states we find the concrete form of $F$ which describes
various types of perfect fluid and dark energy.

\subsection{perfect fluid with a barotropic equation of state}

First of all let us note that one of the simplest and popular model
of the Universe is a homogeneous and isotropic one filled with a
perfect fluid with the energy density $\ve = T_0^0$ and pressure $p
= - T_1^1 = -T_2^2 = -T_3^3$ obeying the barotropic equation of
state
\begin{equation}
p = W \ve, \label{beos}
\end{equation}
where $W$ is a constant. Depending on the value of $W$ \eqref{beos}
describes perfect fluid from phantom to ekpyrotic matter, namely
\begin{subequations}
\label{zeta}
\begin{eqnarray}
W &=& 0, \qquad \qquad \qquad {\rm (dust)},\\
W &=& 1/3, \quad \qquad \qquad{\rm (radiation)},\\
W &\in& (1/3,\,1), \quad \qquad\,\,{\rm (hard\,\,Universe)},\\
W &=& 1, \quad \qquad \quad \qquad {\rm (stiff \,\,matter)},\\
W &\in& (-1/3,\,-1), \quad \,\,\,\,{\rm (quintessence)},\\
W &=& -1, \quad \qquad \quad \quad{\rm (cosmological\,\, constant)},\\
W &<& -1, \quad \qquad \quad \quad{\rm (phantom\,\, matter)},\\
W &>& 1, \quad \qquad \quad \qquad{\rm (ekpyrotic\,\, matter)}.
\end{eqnarray}
\end{subequations}
The barotropic model of perfect fluid is used to study the evolution
of the Universe. Most recently the relation \eqref{beos} is
exploited to generate a quintessence in order to explain the
accelerated expansion of the Universe \cite{chjp,zlatev}.

In order to describe the matter given by \eqref{zeta} with a spinor
field let us now substitute $\ve$ and $p$ with $T_0^0$ and $-T_1^1$,
respectively. Thus, inserting $\ve = T_0^0$ and $p = - T_1^1$ from
\eqref{t00s} and \eqref{t11s} into \eqref{beos} we find
\begin{equation}
S \frac{dF}{dS} - (1+W)F + m W S= 0, \label{eos1s}
\end{equation}
with the solution
\begin{equation}
F = \lambda S^{1+W} + mS. \label{sol1}
\end{equation}
Here $\lambda$ is an integration constant. Inserting \eqref{sol1}
into \eqref{t00s} we find that
\begin{equation}
T_0^0 = - \lambda S^{1+W}. \label{lambda}
\end{equation}
Since energy density should be non-negative, we conclude that
$\lambda$ is a negative constant, i.e., $\lambda = - \nu$, with
$\nu$ being a positive constant. So finally we can write the
components of the energy momentum tensor
\begin{subequations}
\begin{eqnarray}
T_0^0 &=& \nu S^{1+W}, \label{t00sf}\\
T_1^1 = T_2^2 = T_3^3 &=& - \nu W S^{1+W}. \label{t11sf}
\end{eqnarray}
\end{subequations}
As one sees, the energy density $\ve = T_0^0$ is always positive,
while the pressure $p = - T_1^1 = \nu W S^{1+W}$ is positive for $W
> 0$, i.e., for usual fluid and negative for $W < 0$, i.e. for dark
energy.

In account of it the spinor field Lagrangian now reads
\begin{equation}
L_{\rm sp} = \frac{i}{2} \biggl[\bp \gamma^{\mu} \nabla_{\mu} \psi-
\nabla_{\mu} \bar \psi \gamma^{\mu} \psi \biggr] - \nu S^{1+W},
\label{lspin1}
\end{equation}
Thus a massless spinor field with the Lagrangian \eqref{lspin1}
describes perfect fluid from phantom to ekpyrotic matter. Here the
constant of integration $\nu$ can be viewed as constant of
self-coupling. A detailed analysis of this study was given in
\cite{shikin}.

\subsection{Chaplygin gas}

An alternative model for the dark energy density was used by
Kamenshchik {\it et al.} \cite{kamen}, where the authors suggested
the use of some perfect fluid but obeying "exotic" equation of
state. This type of matter is known as {\it Chaplygin gas}. The fate
of density perturbations in a Universe dominated by the Chaplygin
gas, which exhibit negative pressure was studied by Fabris {\it et
al.} \cite{fabris}. Model with Chaplygin gas was also studied in the
Refs. \cite{dev,sen}.

Let us now generate a Chaplygin gas by means of a spinor field. A
Chaplygin gas is usually described by a equation of state
\begin{equation}
p = -A/\ve^\gamma. \label{chap}
\end{equation}
Then in case of a massless spinor field for $F$ one finds
\begin{equation}
\frac{(-F)^\gamma d(-F)}{(-F)^{1+\gamma} - A} = \frac{dS}{S},
\label{eqq}
\end{equation}
with the solution
\begin{equation}
-F = \bigl(A + \lambda S^{1+\gamma}\bigr)^{1/(1+\gamma)}.
\label{chapsp}
\end{equation}
On account of this for the components of energy momentum tensor we
find
\begin{subequations}
\begin{eqnarray}
T_0^0 &=& \bigl(A + \lambda S^{1+\gamma}\bigr)^{1/(1+\gamma)}, \label{edchapsp}\\
T_1^1 = T_2^2 = T_3^3 &=& A/\bigl(A + \lambda
S^{1+\gamma}\bigr)^{\gamma/(1+\gamma)}. \label{prchapsp}
\end{eqnarray}
\end{subequations}
As was expected, we again get positive energy density and negative
pressure.

Thus the spinor field Lagrangian corresponding to a Chaplygin gas
reads
\begin{equation}
L_{\rm sp} = \frac{i}{2} \biggl[\bp \gamma^{\mu} \nabla_{\mu} \psi-
\nabla_{\mu} \bar \psi \gamma^{\mu} \psi \biggr] - \bigl(A + \lambda
S^{1+\gamma}\bigr)^{1/(1+\gamma)}. \label{lspin2}
\end{equation}
Setting $\gamma = 1$ we find the result obtained in \cite{spinpf0}.

\subsection{Modified quintessence}

Finally, we simulate modified quintessence with a nonlinear spinor
field. It should be noted that one of the problems that face models
with dark energy is that of eternal acceleration. In order to get
rid of that problem quintessence with a modified equation of state
was proposed which is given by \cite{pfdenr}
\begin{equation}
p = - W (\ve - \ve_{\rm cr}), \quad W \in (0,\,1), \label{mq}
\end{equation}
Here $\ve_{\rm cr}$ some critical energy density.  Setting $\ve_{\rm
cr} = 0$ one obtains ordinary quintessence. It is well known that as
the Universe expands the (dark) energy density decreases. As a
result, being a linear negative function of energy density, the
corresponding pressure begins to increase. In case of an ordinary
quintessence the pressure is always negative, but for a modified
quintessence as soon as $\ve_{\rm q}$ becomes less than the critical
one, the pressure becomes positive.

Inserting $\ve = T_0^0$ and $p = - T_1^1$ into \eqref{mq} we find
\begin{equation}
F = - \eta S^{1-W} + mS + \frac{W}{1-W}\ve_{\rm cr},  \label{Fmq}
\end{equation}
with $\eta$ being a positive constant. On account of this for the
components of energy momentum tensor we find
\begin{subequations}
\begin{eqnarray}
T_0^0 &=& \eta S^{1-W} - \frac{W}{1-W}\ve_{\rm cr}, \label{edmq}\\
T_1^1 = T_2^2 = T_3^3 &=& \eta  W S^{1-W} - \frac{W}{1-W}\ve_{\rm
cr}. \label{prmq}
\end{eqnarray}
\end{subequations}

We see that a nonlinear spinor field with specific type of
nonlinearity can substitute perfect fluid and dark energy, thus give
rise to a variety of evolution scenario of the Universe.

\section{Cosmological models with a spinor field}

In the previous section we showed that the perfect fluid and the
dark energy can be simulated by a nonlinear spinor field. In the
section II the nonlinearity was the subject to self-action. In
\cite{spinpf0} we have also considered the case when the
nonlinearity was induced by a scalar field. It was also shown the in
our context the results for induced nonlinearity is some special
cases those of self-interaction. Taking it into mind we study the
evolution an Universe filled with a nonlinear spinor field given by
the Lagrangian \eqref{lspin}, with the nonlinear term $F$ is given
by \eqref{sol1}, \eqref{chapsp} and \eqref{Fmq}.

\subsection{Bianchi type-I anisotropic cosmological model}

We consider the anisotropic Universe given by the Bianchi type-I
(BI) space-time
\begin{eqnarray}
ds^2 = dt^2 - a_1^2 dx^2 -  a_2^2 dy^2 - a_3^2 dz^2, \label{BI}
\end{eqnarray}
with $a_i$ being the functions of $t$ only. The Einstein equations
for BI metric read
\begin{subequations}
\label{BIE}
\begin{eqnarray}
\frac{\ddot a_2}{a_2} +\frac{\ddot a_3}{a_3} + \frac{\dot
a_2}{a_2}\frac{\dot a_3}{a_3}&=&  \kappa T_{1}^{1},\label{11}\\
\frac{\ddot a_3}{a_3} +\frac{\ddot a_1}{a_1} + \frac{\dot
a_3}{a_3}\frac{\dot a_1}{a_1}&=& \kappa T_{2}^{2},\label{22}\\
\frac{\ddot a_1}{a_1} +\frac{\ddot a_2}{a_2} + \frac{\dot
a_1}{a_1}\frac{\dot a_2}{a_2}&=&  \kappa T_{3}^{3},\label{33}\\
\frac{\dot a_1}{a_1}\frac{\dot a_2}{a_2} +\frac{\dot
a_2}{a_2}\frac{\dot a_3}{a_3} +\frac{\dot a_3}{a_3}\frac{\dot
a_1}{a_1}&=& \kappa T_{0}^{0}, \label{00}
\end{eqnarray}
\end{subequations}
where dot denotes differentiation by $t$.

From the spinor field equation, it can be shown that \cite{sahaprd}
\begin{equation}
S = \frac{C_0}{\tau}, \label{S}
\end{equation}
where we define
\begin{eqnarray}
\tau =  \sqrt{-g} = a_1 a_2 a_3. \label{taudef}
\end{eqnarray}
For the components of the spinor field we obtain
\begin{eqnarray}
\psi_{1,2}(t) = \frac{C_{1,2}}{\sqrt{\tau}}\,e^{i\int {\cD}
dt},\quad \psi_{3,4}(t) = \frac{C_{3,4}}{\sqrt{\tau}}\,e^{-i\int
{\cD} dt},
\end{eqnarray}
where ${\cD} = dF/dS$. Solving the Einstein equation for the metric
functions one finds \cite{sahaprd}
\begin{eqnarray}
a_i = D_i \tau^{1/3} \exp{\Bigl(X_i \int \frac{dt}{\tau}\Bigr)}
\end{eqnarray}
with the constants $D_i$ and $X_i$ obeying
\begin{equation}
\prod_{i=1}^{3} D_i = 1, \quad     \sum_{i=1}^{3} X_i = 0.
\end{equation}
Thus the components of the spinor field and metric functions are
expressed in terms of $\tau$. From the Einstein equations one finds
the equation for $\tau$ \cite{sahaprd}
\begin{eqnarray}
\frac{\ddot \tau}{\tau}= \frac{3}{2}\kappa
\Bigl(T_{1}^{1}+T_{0}^{0}\Bigr). \label{dtau}
\end{eqnarray}
In case of \eqref{lspin1} on account of \eqref{S} Eq. \eqref{dtau}
takes the form
\begin{equation}
\ddot \tau = (3/2) \kappa \nu C_0^{1+W} (1-W) \tau^{-W}
\end{equation}
with the solution in quadrature
\begin{equation}
\int\frac{d\tau}{\sqrt{3 \kappa \nu C_0^{1+W} \tau^{1-W} + C_1}} = t
+ t_0.
\end{equation}
Here $C_1$ and $t_0$ are the integration constants.

\myfigures{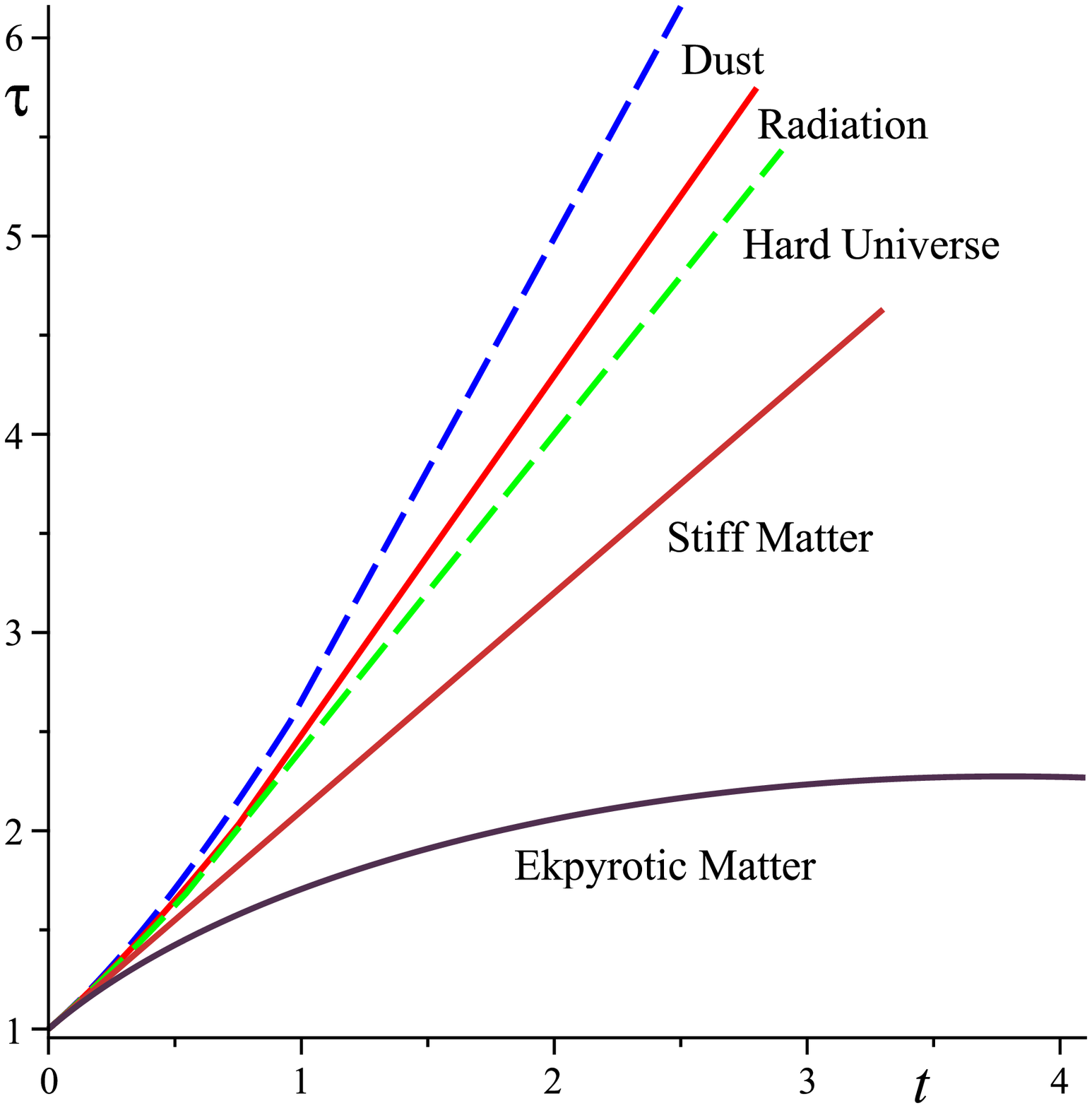}{0.45}{Evolution of the Universe filled with
perfect fluid.} {0.45}{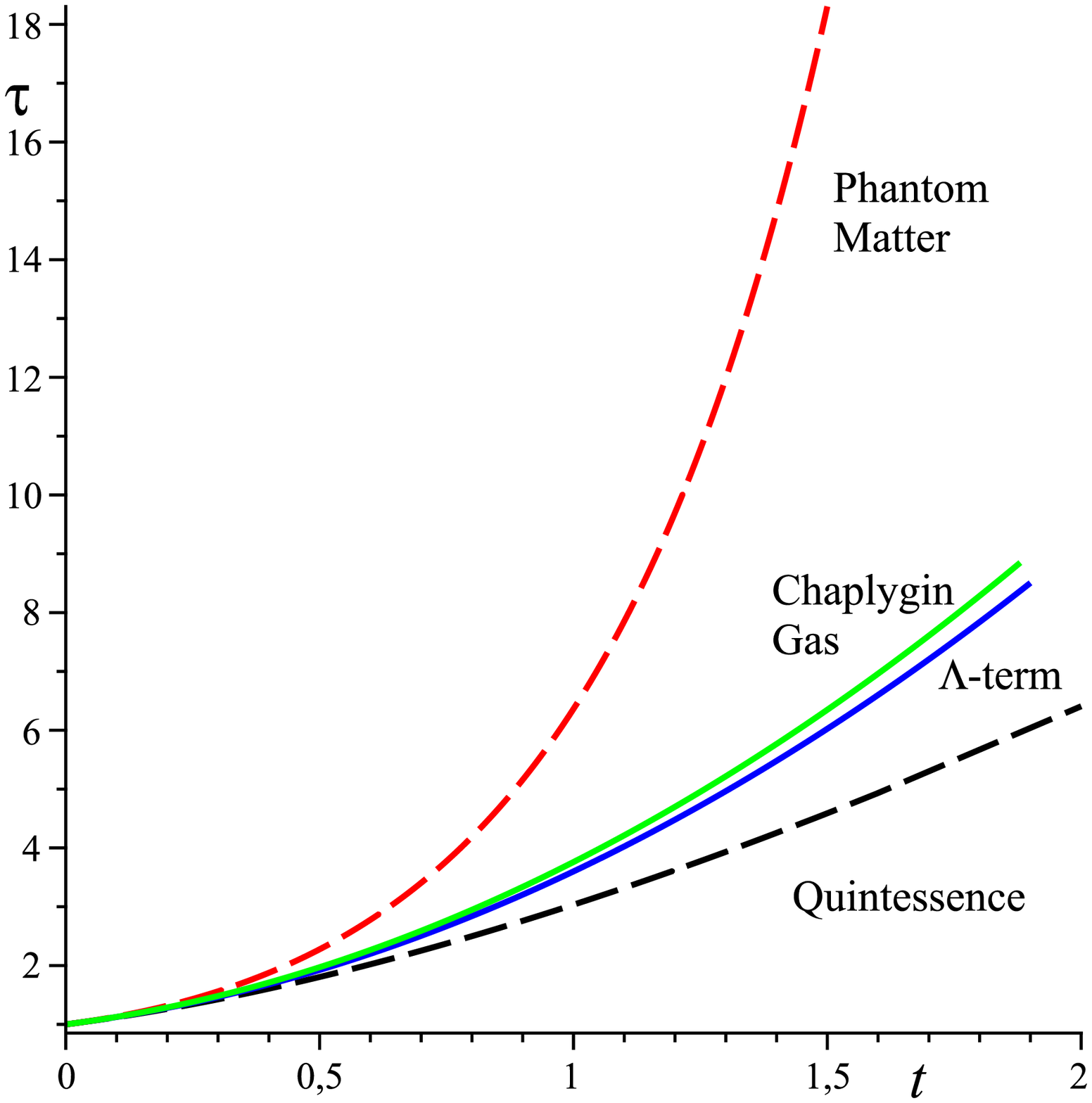}{0.45}{Evolution of the Universe
filled with dark energy.}{0.45}

In the Figs. \ref{spinpf_pf1} and \ref{spinpf_de1} we have plotted
the evolution of the Universe defined by the nonlinear spinor field
corresponding to perfect fluid and dark energy \cite{spinpf1}.

Let us consider the case when the spinor field is given by the
Lagrangian \eqref{lspin2}. The equation for $\tau$ now reads
\begin{equation}
\ddot \tau = (3/2) \kappa \Biggl[ \bigl(A\tau^{1+\gamma} + \lambda
C_0^{1+\gamma}\bigr)^{1/(1+\gamma)} + A
\tau^{1+\gamma}/\bigl(A\tau^{1+\gamma} + \lambda
C_0^{1+\gamma}\bigr)^{\gamma/(1+\gamma)}\Biggr],
\end{equation}
with the solution
\begin{equation}
\int \frac{d \tau}{\sqrt{C_1 + 3 \kappa \tau \bigl(A\tau^{1+\gamma}
+ \lambda C_0^{1+\gamma}\bigr)^{1/(1+\gamma)}}} = t + t_0, \quad C_1
= {\rm const}. \quad t_0 = {\rm const}.
\end{equation}
Inserting $\gamma = 1$ we come to the result obtained in
\cite{chjp}.

Finally we consider the case with modified quintessence. In this
case for $\tau$ we find
\begin{equation}
\ddot \tau = (3/2) \kappa \Bigl[\eta C_0^{1-W} (1+W) \tau^{W} - 2W
\ve_{\rm cr}\tau/(1-W)\Bigr],
\end{equation}
with the solution in quadrature
\begin{equation}
\int \frac{d\tau}{\sqrt{3 \kappa \bigl[\eta C_0^{1-W} \tau^{1+W} -
W\ve_{\rm cr}\tau^2/(1-W)\bigr]  + C_1}} = t + t_0. \label{qdmq}
\end{equation}
Here $C_1$ and $t_0$ are the integration constants. Comparing
\eqref{qdmq} with those with a negative $\Lambda$-term we see that
$\ve_{\rm cr}$ plays the role of a negative cosmological constant.

\myfigures{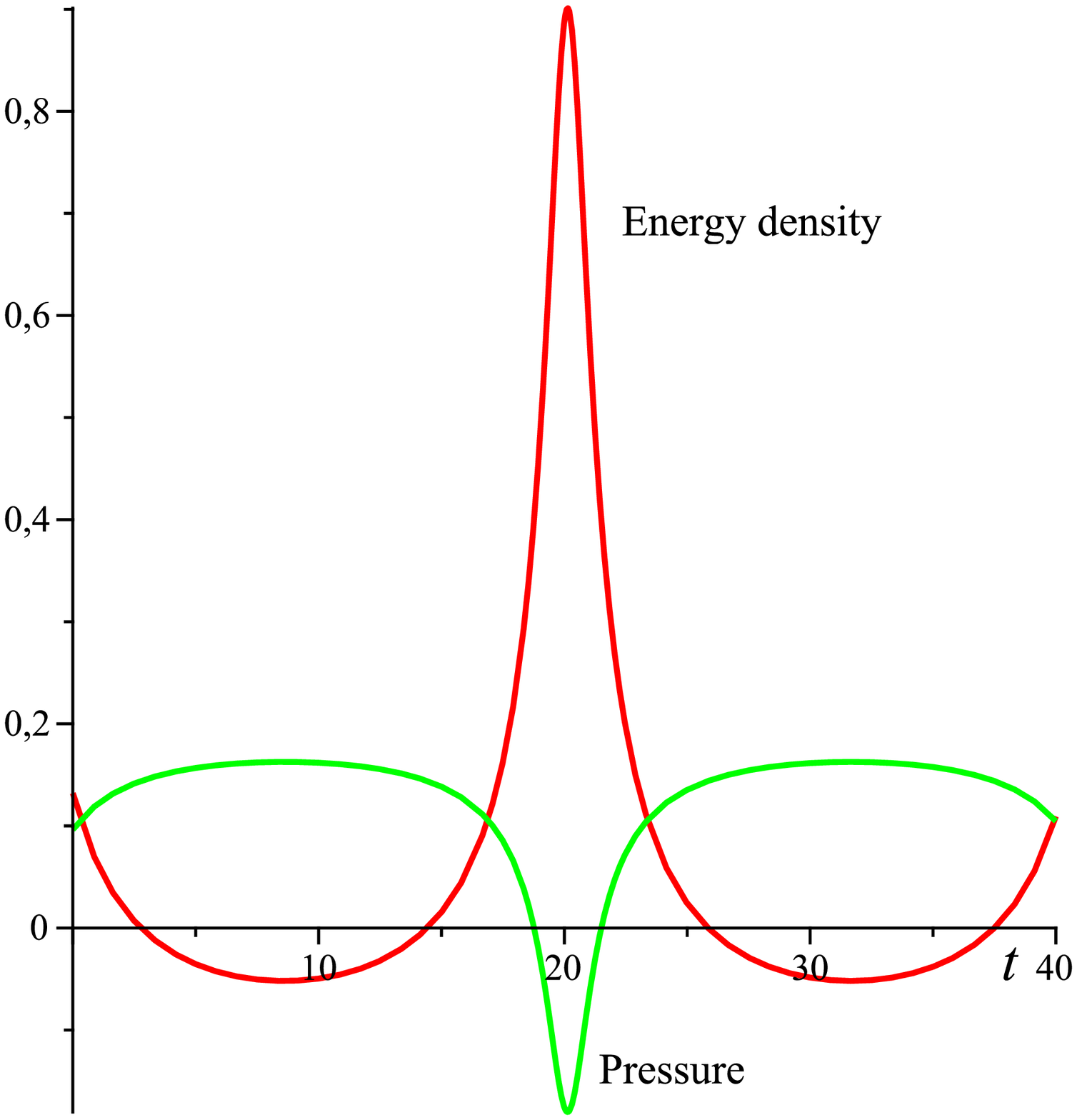}{0.45}{Dynamics of energy density and
pressure for a modified quintessence.}
{0.45}{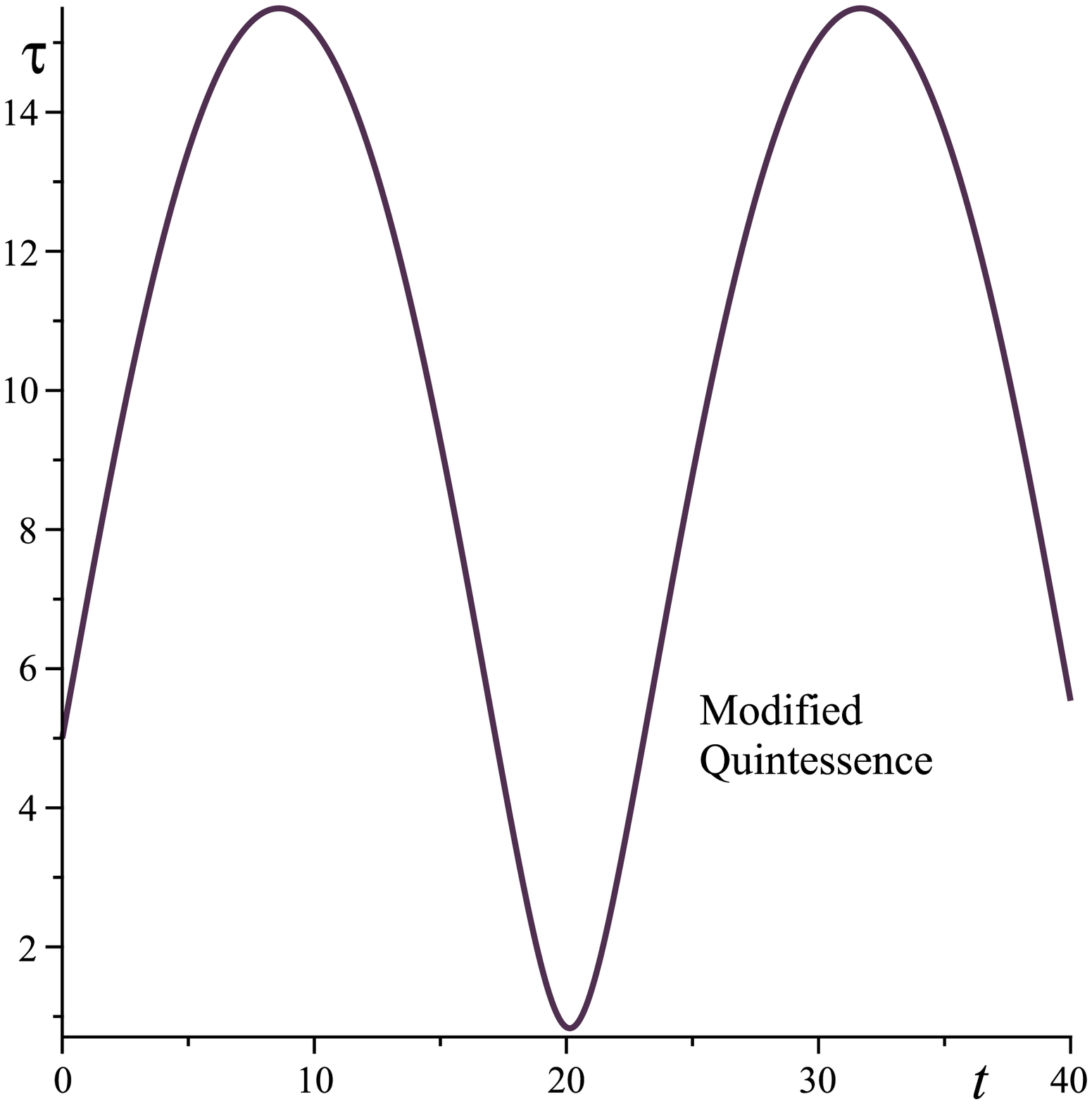}{0.45}{Evolution of the Universe filled with a
modified quintessence.}{0.45}

In the Fig. \ref{spinpf_mqep1} we have illustrated the dynamics of
energy density and pressure of a modified quintessence. In the Fig.
\ref{spinpf_mq1} the evolution of the Universe defined by the
nonlinear spinor field corresponding to a modified quintessence has
been presented. As one sees, in the case considered, acceleration
alternates with declaration. In this case the Universe can be either
singular (that ends in Big Crunch) or regular.

\subsection{FRW cosmological models with a spinor field}

Let us now consider the homogeneous and isotropic FRW cosmological
model with the metric
\begin{eqnarray}
ds^2 = dt^2 - a^2 (dx^2 + dy^2 + dz^2). \label{FRW}
\end{eqnarray}
Corresponding Einstein equations read
\begin{subequations}
\label{EFRW}
\begin{eqnarray}
2 \frac{\ddot a}{a} + \frac{\dot a^2}{a^2}&=& \kappa T_{1}^{1}
\label{FRW11}\\
3\frac{\dot a^2}{a^2}&=& \kappa T_{0}^{0}. \label{FRW00}
\end{eqnarray}
\end{subequations}

From the spinor field equation in this case we find
\begin{equation}
S = \frac{C_0}{a^3}, \quad C_0 = {\rm const.} \label{SFRW}
\end{equation}
The components of the spinor field in this case take the form
\begin{eqnarray}
\psi_{1,2}(t) = \frac{C_{1,2}}{a^{3/2}}\,e^{i\int {\cD} dt},\quad
\psi_{3,4}(t) = \frac{C_{3,4}}{a^{3/2}}\,e^{-i\int {\cD} dt},
\end{eqnarray}
where ${\cD} = dF/dS$.

In order to find the solution that satisfies both \eqref{FRW11} and
\eqref{FRW00} we rewrite \eqref{FRW11} in view of \eqref{FRW00} in
the following form:
\begin{equation}
\ddot a = \frac{\kappa}{6}\Bigl(3 T_1^1 - T_0^0\Bigr) a. \label{dda}
\end{equation}

Further we solve this equation for concrete choice of source field.

Let us consider the case of perfect fluid given by the barotropic
equation of state. In account of \eqref{t00sf}, \eqref{t11sf} and
\eqref{SFRW} \eqref{dda} takes the form
\begin{equation}
\ddot a = \frac{\kappa \nu (1+3W)C_0^{1+W}}{2} a^{-(2+3W)},
\label{ddasf}
\end{equation}
that admits the first integral
\begin{equation}
\dot a^2 =  \frac{\kappa}{3} \nu C_0^{1+W} a^{-(1+3W)} + E_1, \quad
E_1 = {\rm const}. \label{dda1}
\end{equation}

In Fig. \ref{spinpf_pf_FRW} and \ref{spinpf_phan_FRW} we plot the
evolution of the FRW Universe for different values of $W$.

\myfigures{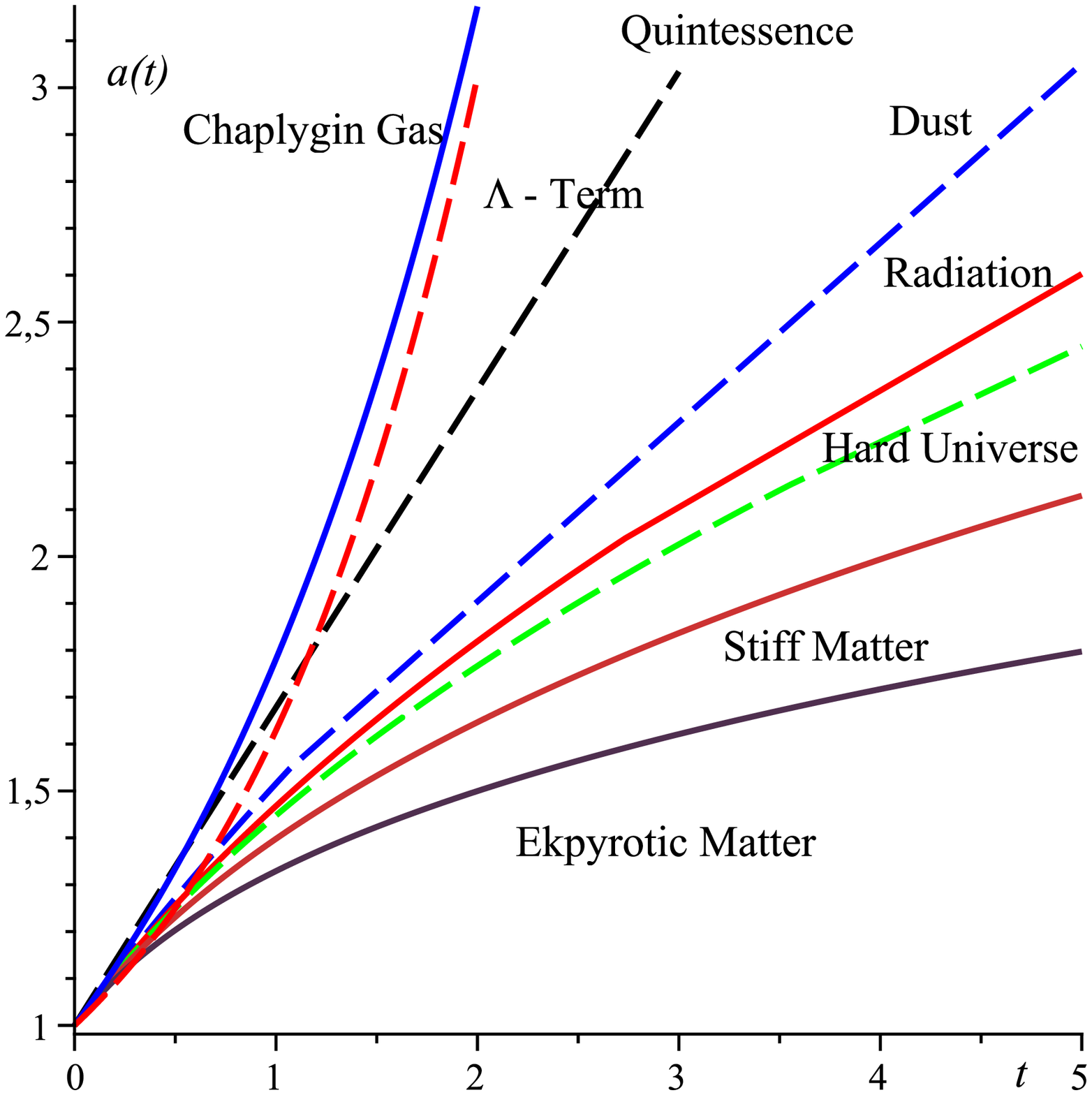}{0.45}{Evolution of the Universe filled
with perfect fluid and dark energy.}
{0.45}{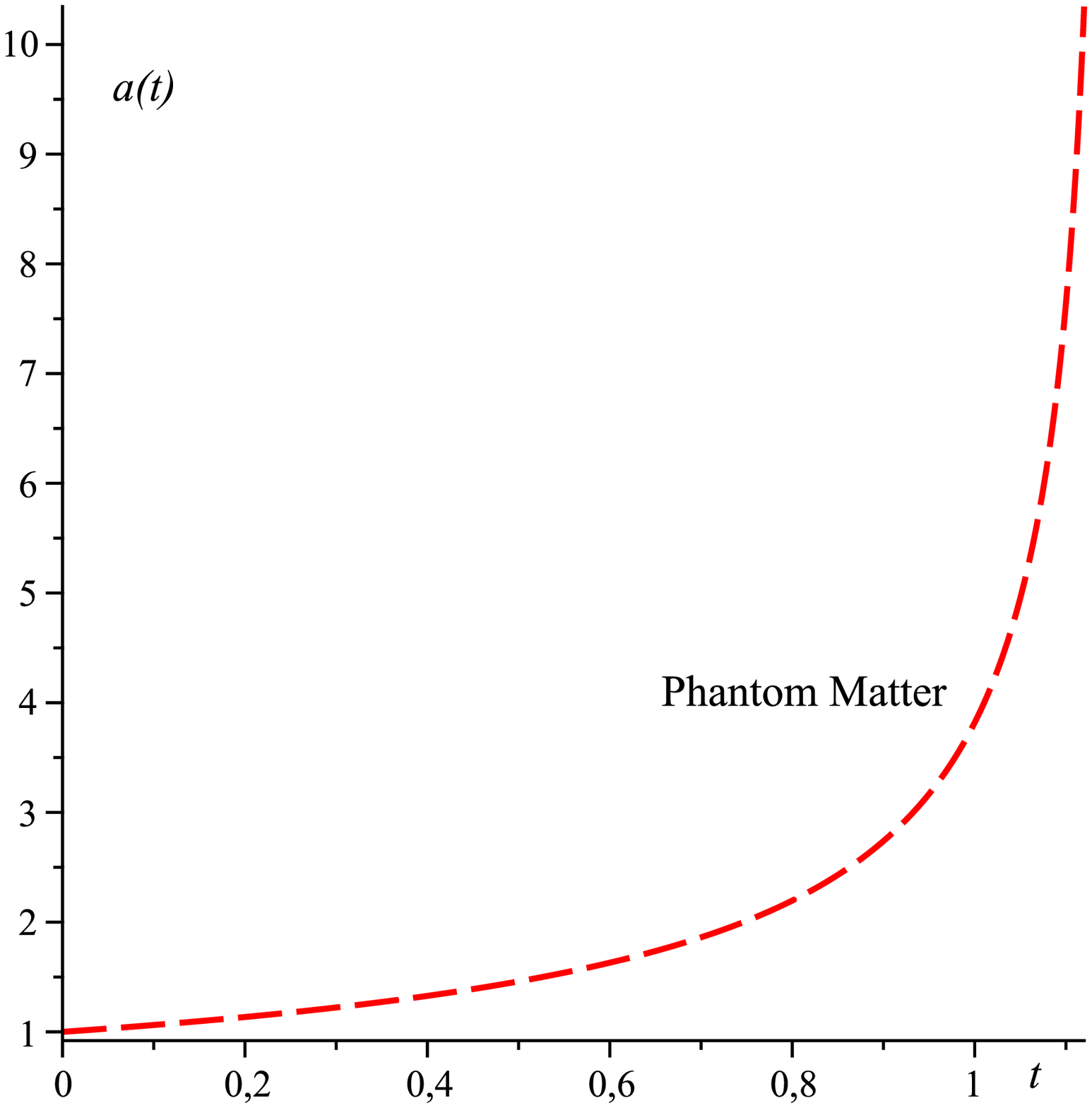}{0.45}{Evolution of the Universe filled with
phantom matter.}{0.45}

As one sees, equation \eqref{dda1} imposes no restriction on the
value of $W$. But it is not the case, when one solves \eqref{FRW00}.
Indeed, inserting $T_0^0$ from \eqref{t00sf} into \eqref{FRW00} one
finds
\begin{equation}
a = (A_1 t + C_1)^{2/3(1+W)}, \label{afrw}
\end{equation}
where $A_1 = (1+W)\sqrt{3 \kappa \nu C_0^{1+W}/4}$ and $C_1 = 3 (1 +
W) C/2$ with $C$ being some arbitrary constant. This solution
identically satisfies the equation \eqref{FRW11}. As one sees, case
with $W = -1$, cannot be realized here. In that case one has to
solve the equation \eqref{FRW00} straight forward. As far as phantom
matter ($W < -1$) is concerned, there occurs some restriction on the
value of $C$, as in this case $A_1$ is negative and for the $C_1$ to
be positive, $C$ should be negative. As one can easily verify, in
case of cosmological constant with $W = -1$ Eqn. \eqref{FRW00} gives
\begin{equation}
a  = a_0 e^{\pm \sqrt{\kappa \nu/3}\,t}. \label{FRWlambda}
\end{equation}

Inserting \eqref{edchapsp} and \eqref{prchapsp} into \eqref{dda} in
case of Chaplygin gas we have the following equation
\begin{equation}
\ddot a  = \frac{\kappa}{6}\frac{2A a^{3(1+\gamma)} - \lambda
C_0^{1+\gamma}}{a^2\Bigl(Aa^{3(1+\gamma)} + \lambda
C_0^{1+\gamma}\Bigr)^{\gamma/(1+\gamma)}}. \label{FRWchap}
\end{equation}
We solve this equation numerically. The corresponding solution has
been illustrated in Fig. \ref{spinpf_pf_FRW}.

Finally we consider the case with modified quintessence. Inserting
\eqref{edmq} and \eqref{prmq} into \eqref{dda} in this case we find
\begin{equation}
\ddot a  = \frac{\kappa}{6}\Bigr[(3W-1)\eta C_0^{1-W} a^{3W-2} -
\frac{2W}{1-W}\ve_{\rm cr} a\Bigr], \label{FRWmq}
\end{equation}
with he solution
\begin{equation}
\dot a^2 =  \frac{\kappa}{3}\Bigl[\nu C_0^{1-W} a^{3W - 1} -
\frac{W}{1-W} \ve_{\rm cr} a^2 + E_2\Bigr], \quad E_2 = {\rm const}.
\label{dda2}
\end{equation}
In Fig. \ref{spinpf_mqep1_FRW}  we plot the dynamics of energy
density and pressure. The Fig. \ref{spinpf_mq_FRW} shows the
evolution of a FRW Universe filled with modified quintessence.

\myfigures{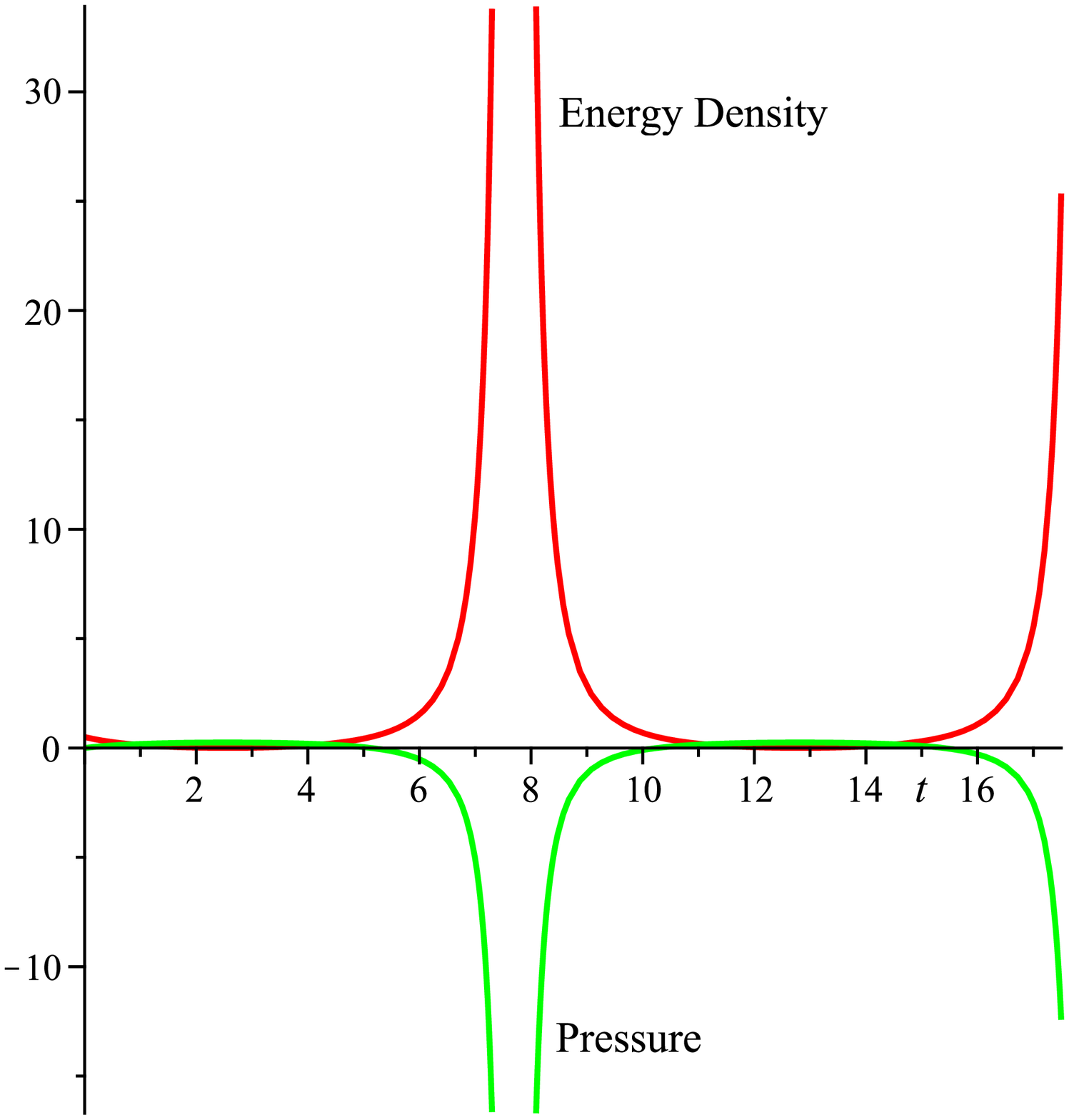}{0.45}{Dynamics of energy density and
pressure for a modified quintessence.}
{0.45}{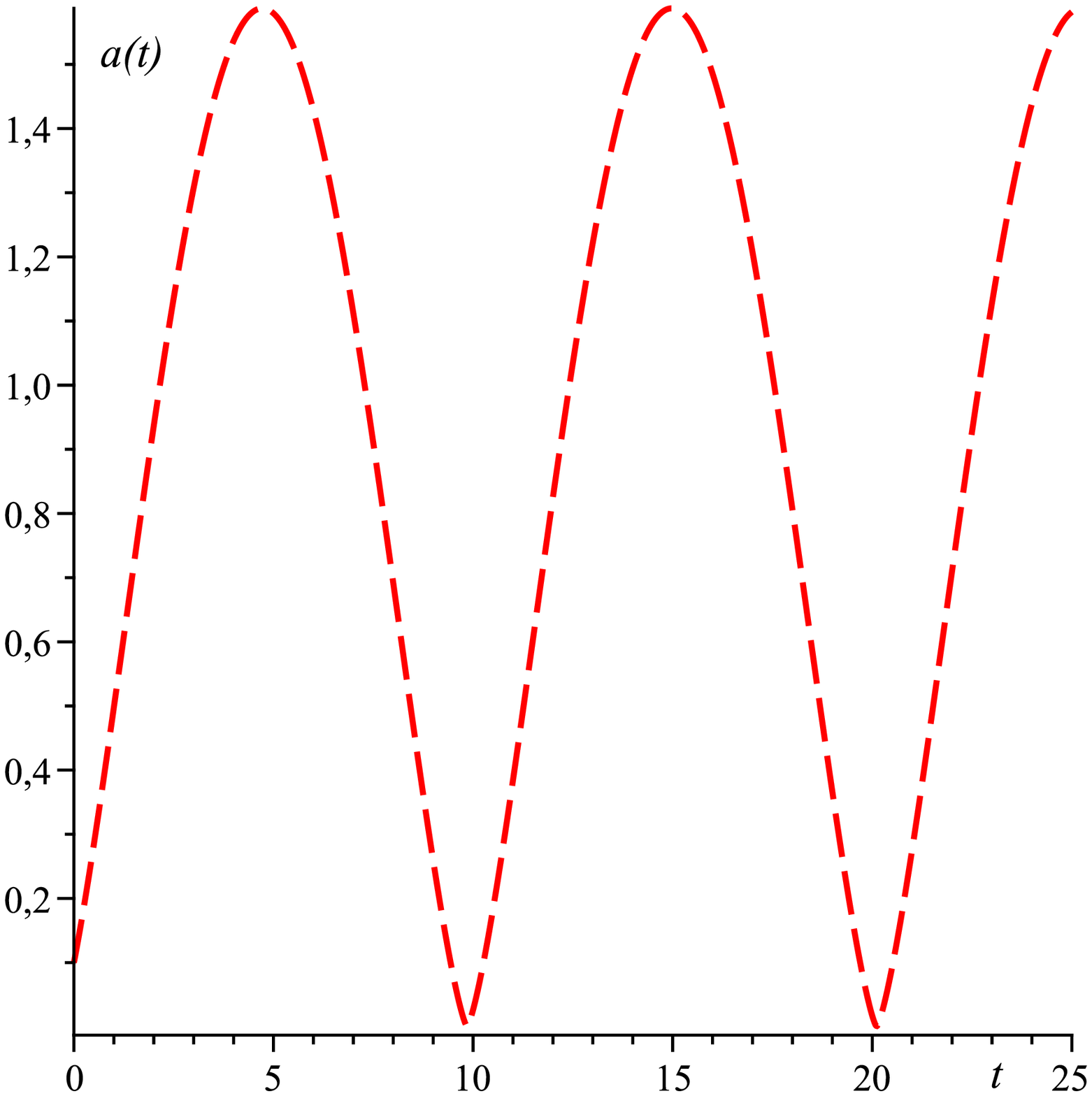}{0.45}{Evolution of the FRW Universe filled
with a modified quintessence.}{0.45}

As one sees, in case of modified quintessence the pressure is sign
alternating. As a result we have a cyclic mode of evolution.

\section{Conclusion}

Within the framework of cosmological gravitational field equivalence
between the perfect fluid (and dark energy) and nonlinear spinor
field has been established. It is shown that different types of dark
energy can be simulated by means of a nonlinear spinor field. Using
the new description of perfect fluid or dark energy evolution of the
Universe has been studied within the scope of a BI and FRW models.

\end{document}